\documentclass[aps,pra,twocolumn,superscriptaddress,nobibnotes,letterpaper]{revtex4-1}

\usepackage[pdftex]{graphicx}
\usepackage{amsmath}
\usepackage[version=4]{mhchem}
\usepackage{verbatim}
\usepackage{color}
\usepackage[version=4]{mhchem}
\usepackage{amsmath}
\usepackage{booktabs}
\usepackage{pifont}
\usepackage{pbox}
\usepackage{verbatim}
\usepackage{natbib}
\usepackage{dashrule}

\usepackage[colorlinks=true, allcolors=blue]{hyperref} 

\usepackage{soul}

\definecolor{chromeyellow}{rgb}{1.0, 0.65, 0.0}

\begin{document}

\title{Fabrication of on-chip probes for double-tip scanning tunneling microscopy}\thanks{This work was published in \href{https://doi.org/10.1038/s41378-020-00209-y}{Microsyst.\ Nanoeng.\ \textbf{6}, 99} (2020).}

\author{Maarten Leeuwenhoek}
\affiliation{Kavli Institute of Nanoscience, Department of Quantum Nanoscience, Delft University of Technology, Lorentzweg 1, 2628CJ Delft, The Netherlands}
\affiliation{Leiden Institute of Physics, Leiden University, Niels Bohrweg 2, 2333CA Leiden, The Netherlands}

\author{Freek Groenewoud}
\affiliation{Leiden Institute of Physics, Leiden University, Niels Bohrweg 2, 2333CA Leiden, The Netherlands}

\author{Kees van Oosten}
\affiliation{Leiden Institute of Physics, Leiden University, Niels Bohrweg 2, 2333CA Leiden, The Netherlands}

\author{Tjerk Benschop}
\affiliation{Leiden Institute of Physics, Leiden University, Niels Bohrweg 2, 2333CA Leiden, The Netherlands}

\author{Milan P. Allan}
\email[]{allan@physics.leidenuniv.nl}
\affiliation{Leiden Institute of Physics, Leiden University, Niels Bohrweg 2, 2333CA Leiden, The Netherlands}

\author{Simon Gr\"{o}blacher}
\email[]{s.groeblacher@tudelft.nl}
\affiliation{Kavli Institute of Nanoscience, Department of Quantum Nanoscience, Delft University of Technology, Lorentzweg 1, 2628CJ Delft, The Netherlands}

\begin{abstract}
	A reduction of the inter-probe distance in multi-probe and double-tip STM down to the nanometer scale has been a longstanding and technically difficult challenge. Recent multi-probe systems have allowed for significant progress by achieving distances of around 30~nm using two individually driven, traditional metal wire tips. For situations where simple alignment and a fixed separation can be advantageous, we here present the fabrication of on-chip double-tip devices that incorporate two mechanically fixed gold tips with a tip separation of only 35~nm. We utilize the excellent mechanical, insulating and dielectric properties of high quality SiN as a base material to realize easy-to-implement, lithographically defined and mechanically stable tips. With their large contact pads and adjustable footprint these novel tips can be easily integrated with most existing commercial combined STM/AFM systems.  
\end{abstract}

\maketitle

\section{Introduction}

Scanning tunneling microscopy using two tips simultaneously in tunneling, also called double-tip STM, often relies on two individually driven metal wire probes brought into close proximity to locally probe resistivity~\cite{thamankar2013low,jaschinsky2008nanoscale,hasegawa2008four} or to access the proposed electron correlations at the nanoscale~\cite{Niu1995,byers1995probing,SettnesPRL,SettnesPRB,leeuwenhoek2020modeling,Ruitenbeek2011,Buttiker1998,Buttiker1999}. Achieving tip separation down to the nanometer scale, a long standing goal in multi-probe STM, has proven challenging and is limited by the radius of curvature of the two tips~\cite{Hasegawa2007b} and requires sophisticated navigation routines~\cite{Kolmer2017, okamoto2001ultrahigh}. Recently multi-probe systems able to achieve a tip separation down to 30~nm have emerged~\cite{Kolmer2017,thamankar2013low} and have resulted in the first double-tip correlation measurements to date~\cite{kolmer2019electronic}. 
      
These experiments are however suffering from complicated alignment procedures and are hence limited to specialized STM setups. Here we continue to build on earlier work~\cite{leeuwenhoek2019nanofabricated} to create a robust and easy to implement on-chip solution where both tips are integrated on a silicon chip. With our approach, given the joined nature of the tips, we can eliminate the need for an additional scanning electron microscope (SEM) column~\cite{Kolmer2017,Jaschinsky2006,yang2016imaging,cherepanov2012ultra,roychowdhury201430,ge2015development} for navigation and make them compatible with ultra-stable compact Pan-type STM heads widely used for single tip experiments~\cite{pan1999}. Moreover, the millimeter scale contact pads and adjustable footprint allows for easy integration in existing and commercially available STM systems. Existing on-chip scanning probes have already contributed in several branches such as parallel AFM~\cite{Vettiger2000}, scanning near field microscopy~\cite{de2013near}, scanning Hall probes~\cite{ge2016nanoscale,mouaziz2006polymer}, scanning SQUID probes~\cite{huber2008gradiometric}, among others. However the development and use of integrated STM tips has been limited~\cite{gurevich2000a,siahaan2015cleaved,xu1995integrated,nagase2008plane}. Recent proof-of-principle experiments have demonstrated that such \emph{single tip} probes can in fact meet the stringent criteria that STM brings~\cite{siahaan2015cleaved} even under ultra-high vacuum and low temperature conditions~\cite{leeuwenhoek2019nanofabricated}.      
  
The main challenge in realizing multi-tip STMs is to minimize the tip-to-tip distance while maintaining the excellent stability required for prolonged in- and out-of-feedback measurements to obtain high quality topographic and spectroscopic data. The double-tip devices presented in this letter build on the recently introduced SiN based smart tip platform~\cite{leeuwenhoek2019nanofabricated} and now incorporate high-resolution Focused Ion Beam (FIB) milling to achieve nanometer tip separation whilst maintaining a rigid connection through a thin silicon nitride (SiN) support. The combination of the mechanical stability provided by the SiN platform with the high resolution milling yields a unique and straightforward approach to the fabrication of scanning probes and their future use in double-tip STM.

\section{Results} 

\begin{figure}[tb]
		\includegraphics[width=1\columnwidth]{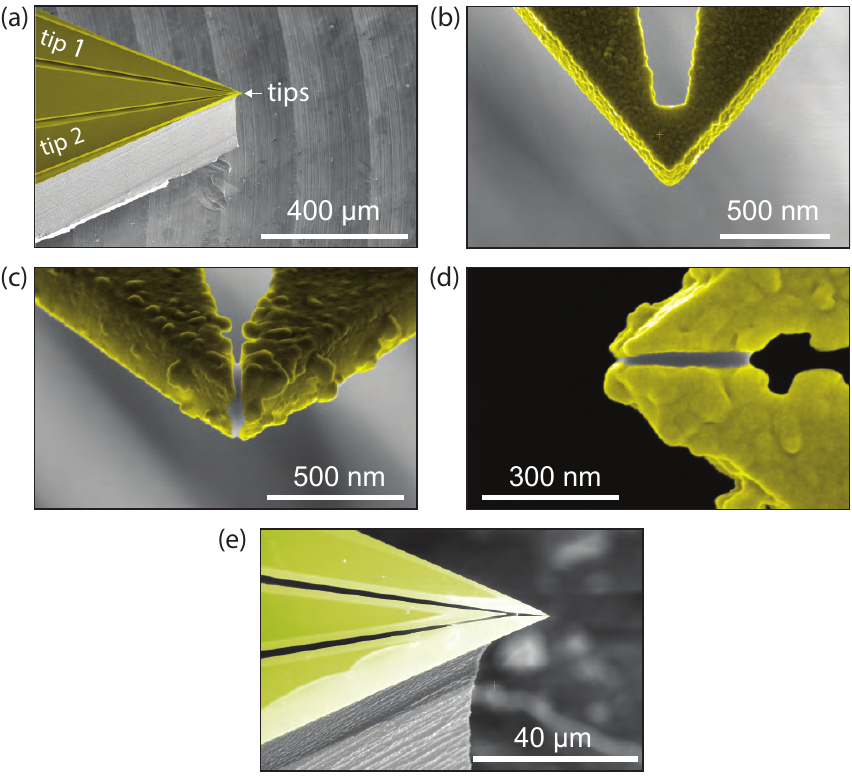}
		\caption{Colored SEM images of several double-tip devices after deposition of \ce{Au} (yellow). (a) Tilted profile of the device showing the chip's profile and the protruding tips. (b) Tip apex before FIB milling. (c) Tilted front view of the two tips after FIB milling. The two tips originate from a 55~nm gold layer and the gap between the tips is around 35~nm and hence the total distance between the points that are most likely to tunnel is approximately 50~nm. The required 3D alignment can be directly estimated from these tilted images. (d) Top view of the same device as (c) showing clear protrusion of the metal film beyond the SiN. (e) Side view of the device shown in (c) and (d).}
		\label{fig:DTfinal}
\end{figure}

Fig.~\ref{fig:DTfinal} gives an overview of several devices we fabricated using the methods described above. Fig.~\ref{fig:DTfinal}a shows an overview of the chip, where the three contact pads separated by trenches can clearly be identified. The narrow bright lines along the pads are the regions consisting of undercut SiN covered in gold like the rest of the chip. The angled view clearly shows the sidewalls of the chip and the overhanging tips where the contact pads meet. Higher magnification images provide a more detailed view of the apices of the tips, where Fig.~\ref{fig:DTfinal}b is taken before the FIB step. Such a device can act as single tip device or potentially be used to perform tip preparation through heating by running a current between the contacts, as suggested by~\cite{ciftci2019polymer}. 

Fig.~\ref{fig:DTfinal}c,d show the apices of two tips from the front under a 52$^{\circ}$ angle and from top, respectively. They allow us to clearly see the metal film coverage on the sides of the SiN layer as well as the separation between the tips which is approx.\ 35~nm edge-to-edge. The relatively thick \ce{Au} film becomes grainy, especially near the apex of the SiN, while it is smoother on the pads. Even though the coarseness of the metal makes the exact location of the tip more uncertain, the metal consistently covers the full apices (Fig.~\ref{fig:DTfinal}d). For this particular device the FIB milling depth is set to 300~nm to ensure a cut through all the metal on the side of the SiN.

\section{Discussion}

The general fabrication process we present follows the SiN based smart-tips introduced earlier for single tip devices~\cite{leeuwenhoek2019nanofabricated}. Here, we continue to build on this platform as we extend the fabrication procedure to create double-tip devices by incorporating several new techniques. The high quality SiN that provides the base for the tips plays a central role in the design and fabrication of the devices. First, their excellent insulating, dielectric and mechanical properties provides us with the opportunity to keep the two tips mechanically attached while electrically separated. Second, the SiN also allows to proceed with a single lithography step that incorporates the additional complexity that the multiple tips and contact pads bring. By transferring the shape of both tips and their contact pads into the SiN we can later use an isotropic silicon etch to create trenches that electrically separate the tips even after deposition of the metal as we will show below.

\subsection{Pattern design}

\begin{figure}[t]
	\includegraphics[width=1\columnwidth]{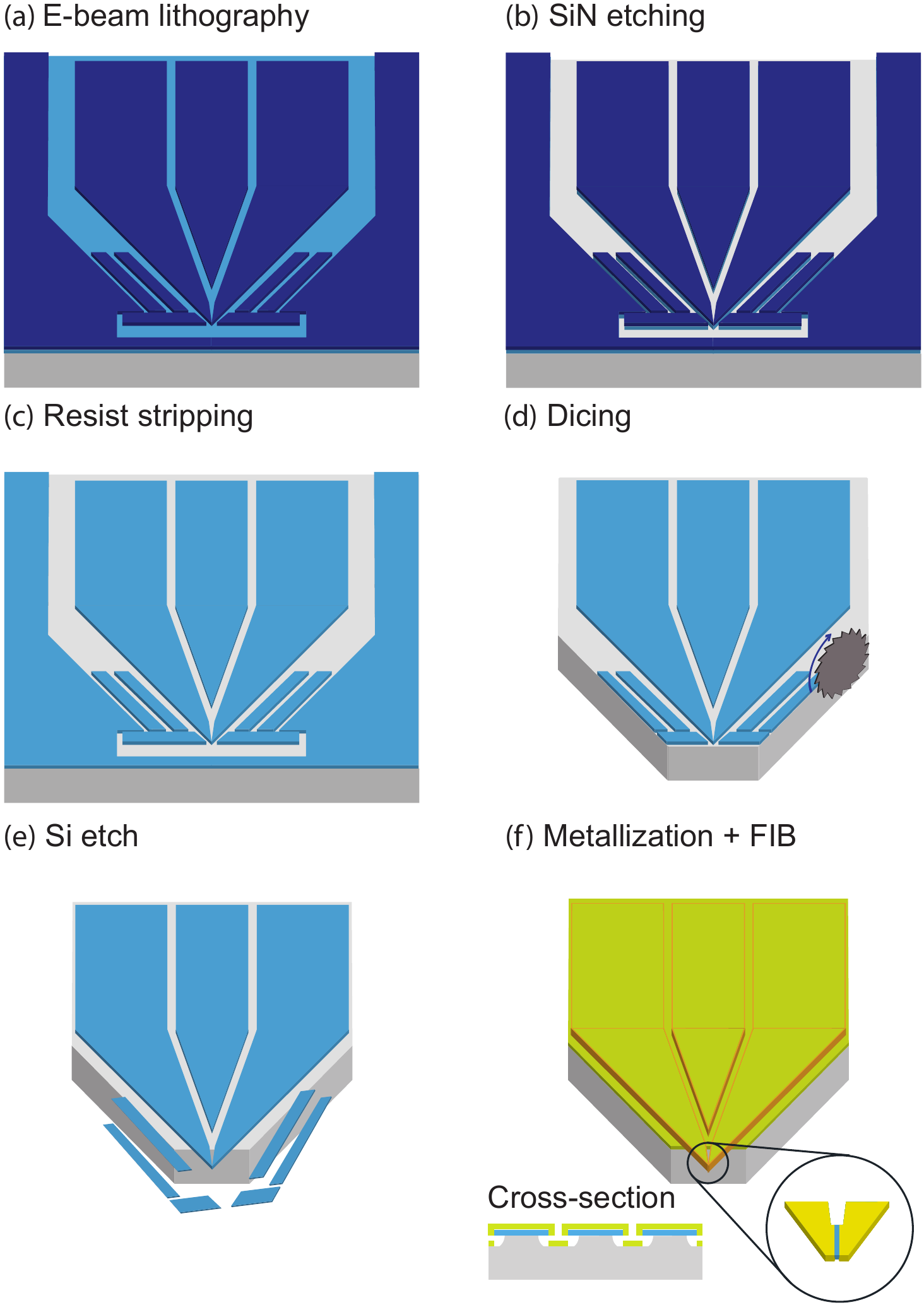}
	\caption{Fabrication procedure for double-tip devices. (a) The resist layer (purple) on the SiN (blue) covered Si(100) chip (gray) is patterned by e-beam lithography, creating a large opening in front of the tips and around the contact pads, as well as the lines that encircle the shields and the tips. (b) The pattern is then transferred into the SiN. (c) The resist is cleaned thoroughly and a fresh layer of photoresist is applied to protect against debris from the dicing (not displayed for clarity). (d) The chip is then diced (more details can be found in Fig.~\ref{fig:dice}). (e) After removal of the protective resist, the chip is undercut using a isotropic Si dry etch, removing the Si substrate primarily from the sidewalls causing the shields to drop off, leaving the tips freestanding. (f) We evaporate a \ce{Au} layer on top of a \ce{Cr} adhesion layer followed by focused ion beam milling to separate the two tips (inset right). The electrical isolation of the contact pad is illustrated in the cross-section inset.}
	\label{fig:smarttipsteps}
\end{figure}

A schematic overview of the full fabrication procedure is shown in Fig.~\ref{fig:smarttipsteps}. We start with a 200~nm thick layer of high stress, low-pressure chemical vapor deposition (LPCVD) silicon nitride on both sides of a 200~$\mu$m thick Si(100) chip and we spincoat a 550~nm thick layer of ARP-6200.13 resist on top for electron beam lithography. The pattern is created by a single exposure but consists of two parts. The first is the tip shape itself with the surrounding shields, which we include to minimize the overhang of the SiN surrounding the tip after the Si etch as described below:\ by covering these areas with the shields the isotropic etch effectively only affects the sidewalls and does not create a significant undercut from the top. The second part consists of the two contact pads and a center pad that is used to reduce the exposed area to shorten the exposure time but also to minimize proximity effects of the exposure near the two tips. For future applications it can also act as a third contact pad (combined AFM/STM systems often include three electrical contacts). The large areas are exposed by a 40~nm electron beam with a beam step size of 20~nm and a dose of 400~$\mu$C/cm$^2$ and the small structures by a smaller beam of 18~nm together with a step size of 2.5~nm and a dose of 320~$\mu$C/cm$^2$, both with a 100~kV beam.
Note that the narrow trenches between the tips/contacts could lead to electrical shorts formed by accidental leftover traces of resist inside the trenches or small pieces of dirt connecting the pads after deposition of the metal. Therefore we enlarge the spacing between the tips and pads as we move away from the apices of the two tips that are still connected at this stage.

We then develop the chip in Pentyl Acetate (1~min) and MIBK:IPA 1:1 (1~min) followed by an IPA rinse (1 min), resulting in Fig.~\ref{fig:smarttipsteps}a. The pattern, consisting of the tip shape and six shields, is transferred into the SiN layer using a \ce{CHF3} etch for 5~min (Fig.~\ref{fig:smarttipsteps}b). Directly after we start the removal of the resist by exposing the chip to a \ce{O2} plasma for 10~min. We then continue cleaning the chip by successive immersions into N-N-Dimethylformamide (DMF) for 10~min, Positive Resist Stripper (PRS) for 10~min, followed by a boiling piranha solution at 135$^{\circ}$C for 8~min to remove all traces of resist and other organic contamination (Fig.~\ref{fig:smarttipsteps}c).

\subsection{Dicing}
\label{Dicing}

To bring the tip to the edge of the chip we proceed with dicing (Fig.~\ref{fig:smarttipsteps}d). Here we prefer dicing over a through-wafer Deep Reactive Ion Etch (DRIE) since it allows to more easily explore different designs with each new iteration and reduces the number of fabrication steps. We would however like to stress that our method is fully compatible with full wafer processing as we discuss later.

\begin{figure}[t]
	\begin{center}
		\includegraphics[width=1\columnwidth]{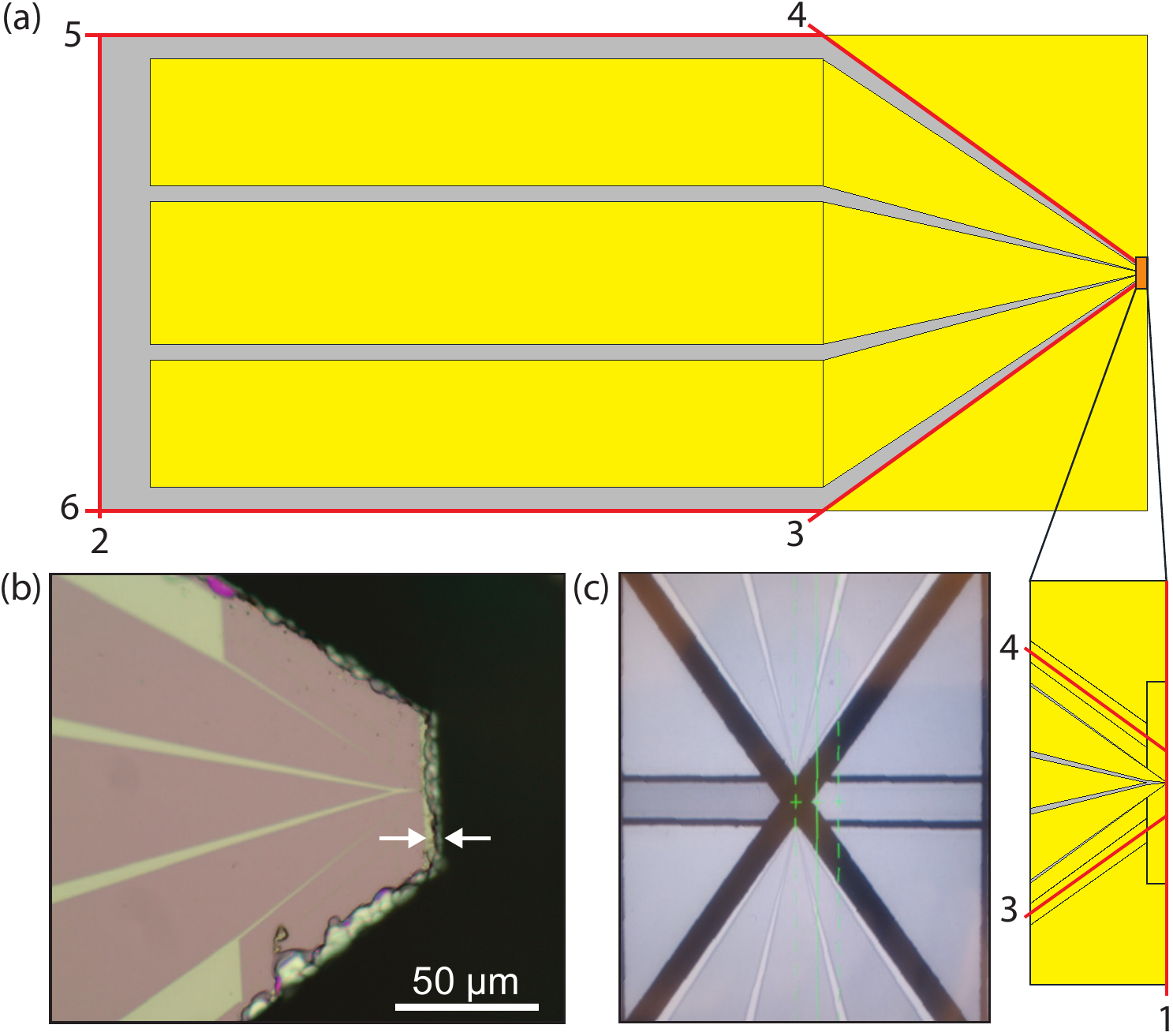}
		\caption{(a) Double-tip pattern design with dice lines shown in red. The gray areas indicate the Si, while the yellow parts are covered with SiN. The small features, such as the tips and the surrounding shields are highlighted in the inset on the bottom right. (b) Optical microscope image of a chip after dicing. The silicon (gray) left after the dicing extends up to 6~$\mu$m beyond the SiN (pink). (c) Screen capture of a chip containing two opposing smart tips after dicing as seen through the microscope of the dicer. For illustrative purposes the image is taken on low magnification, the alignment of the blade is done at high magnification. The black areas are diced.}
		\label{fig:dice}
	\end{center}
\end{figure}

Before the dicing step we first protect the chip surface against any residual debris by applying a new layer of photoresist. From a typical 10x10~mm$^2$ chip, we cut two smart-tips to ensure that the small features of the tip are in the center of the chip to for optimal resist conditions for the EBL (Fig.~\ref{fig:dice}c). Successful dicing results in (i) smooth sidewalls, such that the overhanging tip will be the most protruded feature, (ii) minimal chipping of the Si and, (iii) good alignment accuracy. 

The first requirement we obtain by choosing the optimal blade and settings for the dicer (Disco dicer DAD 3220). We found the Disco ZH05-SD2000-N1-90 blade at a forward speed of 3~mm/s to yield optimal results in terms of chipping and sidewall smoothness. The limited residual roughness on the sidewall is further smoothed out by the isotropic Si etch described in the next section.

To align the dice consistently down to a few microns from the patterned tip, we need to minimize the amount of chipping of the Si along the dice line (Fig.~\ref{fig:dice}b). The blade is therefore typically dressed -- re-sharpened by cutting into a special substrate and increasing the exposure of the diamonds in the blade -- for each new chip (two tips) and we perform the most critical dice (number 1 in Fig.~\ref{fig:dice}a) first as the chipping increases with each dice.

Last, to align accurately to the patterned tips we calibrate the width of the blade accurately by a so called hairline adjustment.

\subsection{Suspending SiN tip}

First we wash away the residue created by the dicing with the removal of the protective photoresist layer and then the chip is cleaned again with a 10~min \ce{O2} plasma. After these cleaning steps, we isotropically remove part of the Si substrate by exposure to a \ce{SF6} plasma for 3-4~min inside an inductively coupled plasma (ICP) etcher. Reactive ion etching is often a combination of chemical reactions, the \ce{F^-} ions reacting with the Si, and ion bombardment, where the ions are accelerated by the bias into the sample hence removing material. Here, however, to prevent any anisotropy in the etch we do apply a bias except for a short ignition pulse $<$1~s to ignite the plasma, resulting in a predominantly isotropic Si etch~\cite{norte2018platform}. To increase the selectivity of the SiN over the Si the chip is cooled to $-50^{\circ}$C.

\subsection{Metallization}

After the overhang is created we proceed to deposit the metal film. While our previous work~\cite{leeuwenhoek2019nanofabricated} relied on sputtered Au films to ensure maximum coverage of the SiN sidewalls, here we use an electron beam evaporator to deposit a 3-5~nm \ce{Cr} adhesion layer and a 45-60~nm thick \ce{Au} film. The evaporation of Au films is more directional, but should in combination with a \ce{Cr} layer underneath show better adhesion to the SiN. To counter the effects of the directionality of evaporation over sputter deposition we place the chip under an angle, typically around 40$^{\circ}$ and rotate it at 10~rpm to obtain the sidewall coverage.

\subsection{Optimizations and challenges}
\label{Challenges}

\begin{figure}[t]
		\includegraphics[width=1\columnwidth]{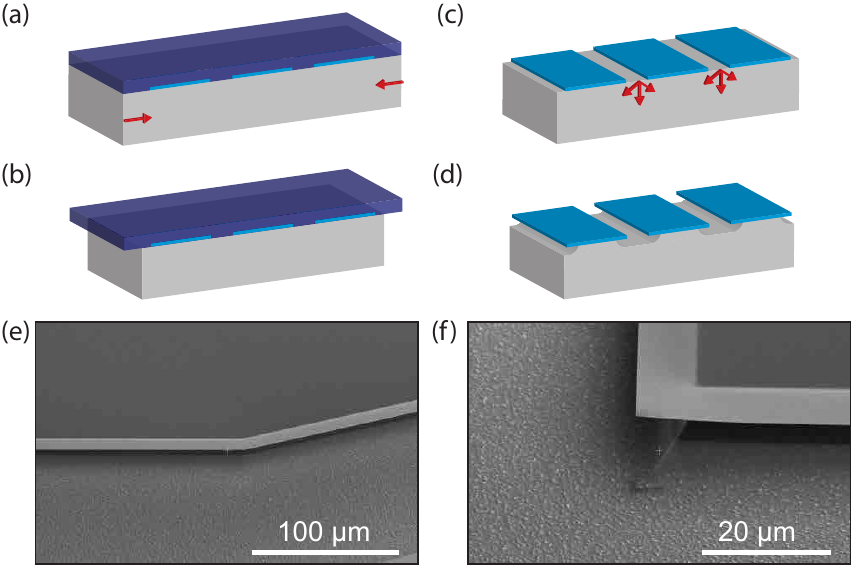}
		\caption{Procedure for minimizing trench depth and width between the contact pads (blue). A layer of photoresist (purple) covers the full chip (a) and we etch only the sides of the chip creating the tip overhang (b). The resist is then removed (c) and an additional very brief isotropic Si etch is used to create the trenches between the pads for electrical separation (d). SEM images of the finished device showing the undercut along a trench (e) and at a corner (f).}
		\label{fig:Undercut}
\end{figure}

Before demonstrating the final devices we introduce an optional optimization:\ The overhangs created on all edges, if they break, can cause shorts to the lower lying metal on the Si especially during mounting in the STM tip holder. To date, both the protrusion of the tips and the overhanging sides of the contact pads have been created with the same isotropic Si etch. We do this after dicing and subsequent removal of the protective photoresist with acetone and IPA. However, we can also choose to leave the resist used during dicing on and perform the etch, as is illustrated in Fig.~\ref{fig:Undercut}. Now only the sidewalls of the chip retract while the top layer is protected, therefore we create the protrusion of the tip, but no trenches between the contact pads. The latter, however, is needed to create electrical separation of the pads. Once the tip overhangs and (almost) all shields are now held by the resist, we spray acetone and IPA to clean off the resist in the direction away from the tip to prevent the shields from landing on top. Finally we can perform a very short additional isotropic etch to form the contact pads. The result is smaller overhanging regions around the sides of the contact pads. How much smaller is determined by how short the last etch is.

\subsection{Focused Ion Beam milling}

The minimum tip-to-tip distance previously achieved by having two separate SiN tips is 45~nm for a test device using 50~nm SiN and (sputtered) 20~nm \ce{Au}~\cite{leeuwenhoek2019nanofabricated}. Here we work with a thicker more reliable SiN film and \ce{Au} layer. This also enlarges the radii of curvatures of the two tips and thereby limits the tip-to-tip distance to more than twice that radius, a similar limitation to bringing two individual tips into close proximity. Moving towards 50~nm thick SiN and a very thin film helps to reduce tip separation but brings uncertainties. First, depending on the length of the overhanging tips, the thin SiN tends to bend upwards due to the intrinsic stress of the film. While this bend is not be a problem in principle, it might not always be equal for both tips when not connected. We therefore decided to use the thicker 200~nm SiN in combination with slightly thicker evaporated Au films.

As a result of the excellent control we have over the shape of the tips, we have created devices where the two apices are joined together in both the SiN and Au film, and separate the tips using focused ion beam (FIB) milling. The aim is to mill only the metal layer and keep the SiN attached. A priori this method present several advantages:\ (i) FIB milling has the ability to achieve higher resolution than EBL~\cite{volkert2007focused} especially when combined with a 200~nm deep etch through the SiN, (ii) tip separation is not determined by the film thickness, (iii) the mechanical connection formed by the SiN eliminates possible height or tip-to-tip distance fluctuations, contrary to earlier reported proof-of-principle devices~\cite{leeuwenhoek2019nanofabricated}, and (iv) improves the overall (macroscopic) stability of the tips. The dielectric breakdown voltage that can result in a short-circuit between the two tips should occur well above 5~V for these structures and will thereby not hinder normal STM operation. Another possibility for the separation of the metal we leave completely unexplored, is the use of electromigration to create the tip separation to possibly even smaller distance down to 5~nm~\cite{heersche2007situ}, although sufficient control over where the separation occurs might be challenging. 

A well known side effect of FIB milling is the addition of \ce{Ga+} ions a few nanometers into the exposed top layer of the material. The impact of the ions can result in lattice defects, incorporated Ga, and heat~\cite{volkert2007focused}. The range of added \ce{Ga+} ions can vary between 10-100~nm deep and 5-50~nm wide. The \ce{Ga+} ions at the \ce{Au} tip surface may be cause for concern if they indeed have an effect on the quality of the STM tips or their electronic properties, however previous STM experiments have used tips modified by FIB before~\cite{schmucker2012field,Kolmer2017}. The difference here is that redeposition of the insulating SiN can occur. However, a short immersion in an HF solution or exposure to vapor HF~\cite{williams2003etch} can be used to remove a few nanometers of SiN without damaging the overall structure of the tips. Furthermore, a common in-situ tip preparation technique called mechanical annealing~\cite{castellanos2012highly,Tewari2017a} allows us to pick up clean gold from a Au(111) surface and may assist in attaining a clean STM tip and will assure tip sharpness. Variation of the tip-to-tip distance due to the graininess of the metal film may still occur, but can be readily improved by optimization of the evaporation temperature, moving to thinner, less grainy films such as \ce{AuPd} and through additional shaping of the tip using FIB.

\subsection{Tip characterization}

To validate the fabrication method described in the paper, we test the device in a modified scanning tunneling microscope (STM) manufactured by RHK.

\begin{figure}[t]
		\includegraphics[width=1\columnwidth]{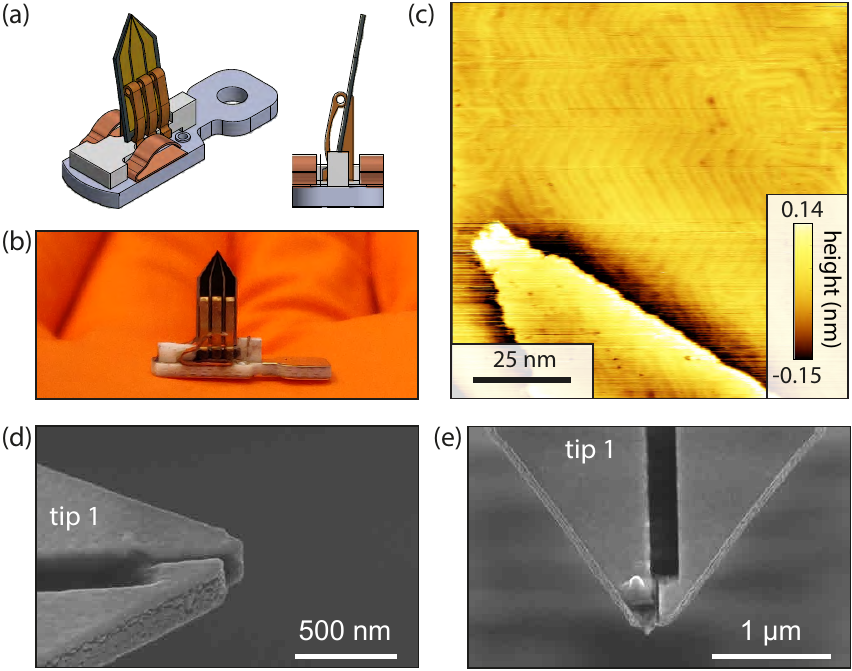}
		\caption{Testing a fabricated device on a Au(111) surface. (a) Schematic drawing of a device holder including the double-tip device. (b) Photograph. (c) Topographic image of a gold on mica sample using our tip for STM imaging taken with setup conditions $V_b=0.75$~V, $I=50$~pA. (d), (e) SEM micrograph of the device used to measure (c) before and after applying a voltage of $>$5V between the two tips.}
		\label{fig:characterization}
\end{figure}

We built a custom holder for the device (Fig.~\ref{fig:characterization}a,b) consisting of three phosphor bronze contact clamps attached to an \ce{Al2O3} block on a base plate made out of a special printed circuit board equipped with vias to contact the STM. The device is then inserted into the holder and held by the contact springs. We insert the holders into our cryogenic ultra-high vacuum STM.

For the imaging and the breakdown experiments we prepare a gold film with three cycles of \ce{Ar+} sputtering (0.75~kV, 3~mA at $5\times10^{-5}$~mbar for 15~min) and subsequent annealing around 390$^{\circ}$C for 30~min under UHV conditions.  We then scan the gold on mica sample. Figure~\ref{fig:characterization}c shows a topographic image from the Au(111) surface exhibiting both step edges and the typical herringbone reconstruction. The gold surface is commonly used to assess tip quality and stability, however, since it is metallic, gating effects from the second probe are not expected as the potential drop occurs completely  between tips and sample. Finally, we investigate the breakdown voltage between the tips by applying voltages of 1~V, 2.5~V, 5~V, and 7.5~V between the tips. The voltage is increased with the tip retracted, and the device is then tested on the Au(111) surface with the voltage fixed. The device continues to function up to 5~V applied between tips, but breaks down at 7.5~V, as confirmed by SEM images (see Fig.~\ref{fig:characterization}d,e).

\vspace{1cm}

\section{Methods} 

\subsubsection{Dicing}

The full dicing procedure works as follows:\ First we dress the blade, load the chip and proceed to make a first cut and perform the hairline adjustment. Then, with a sequence of individually aligned dices as indicated in Fig.~\ref{fig:dice}a,c we cut out the chip shape. Line 1 is the one close to the tips and is therefore performed first. Line 2 follows keeping the same alignment, that way we ensure that, when the chip stands upright, the bottom is exactly parallel to the top where the tips protrudes. The cross lines 3 and 4 run through the shields typically around 20~$\mu$m away from the tip on each side (Fig.~\ref{fig:dice}b). Finally we dice lines 5 and 6 parallel to each other, these set the width of the chip to 3~mm.

\subsubsection{Focused Ion Beam milling}

For the milling procedure we load the chips on a piece of carbon tape into the FIB system with the sample holder firmly screwed in to prevent tiny displacements during tilting. We adjust the focus of the electron beam repeatedly as we move the stage to its working distance. We also correct for stigmatization and make sure the electron beam is not shifted with respect to the ion beam for the alignment procedure. Then we pick a reference point at high magnification and tilt the sample 52$^{\circ}$ towards the ion beam while imaging with the electron beam. As soon as we image with the ion beam it will start milling, therefore we first find an area close to the tip but keep it outside the field of view. We can set up such that the field of views of the electron and ion beam are the same and the magnifications are linked. Before we mill, we focus the ion beam at 100,000x magnification. Using the electron beam we navigate to the tip, take a single scan with the ion beam and align the milling path. After the alignment we typically take a second scan to check and compensate for a slight drift in the image. Then we perform the milling over a length of 1-2~$\mu$m in less than a second using an optimized recipe for \ce{Au} with a milling depth between 200-300~nm.

\section{Conclusion}
\label{Conclusions}

We have demonstrated a new fabrication procedure for integrated double-tip devices with a tip separation of around 35~nm based on our highly flexible SiN based smart tip platform. By etching the contact pads and two tip geometry into the SiN layer, we can create double-tip devices with an easy to implement process. The excellent mechanical and insulating properties of the SiN allow us to keep the tips attached via the SiN and utilize high resolution FIB milling of the metal layer to separate the two tips. This leads to increased mechanical stability, better alignment of the tips with respect to each other and remove the need for any electromechanical actuation~\cite{martin2009nanoelectromechanical}. 

Integration of these devices in existing commercial STM systems is straightforward when multiple tip contacts are available. The chips are only 200~$\mu m$ thick, can be made in various sizes and include large contact pads, making them well suited for the limited space available inside a STM head. Another benefit towards routine use of these tips is the ability to upscale the production of the devices to full wafer processing. By replacing the dicing of individual chips with a through-wafer Bosch etch, one could make a large number of chips from a single 4" wafer. For this we would use a \ce{SiO2} hard mask combined with electron beam or optical lithography to align the chip pattern to the underlying SiN tips/contacts with possibly even better accuracy compared to dicing. Once optimized, this will yield faster, larger and more consistent production.

A fully functioning double-tip STM based on the tips presented here differs significantly from multi-probe system using metal wire tips. Here the tips are joined at a fixed distance from each other on the chip that moves via a single positioning system piezo tube. Consequently, the control over the two tips comes with unique challenges. To get both tips into tunneling will require a tilt stages to compensate for the slight difference in length between the tips and sample tilt. Based on the measurement of 4 different double-tips, we conclude that the typical angles between tips in various devices range from 4 to 15 degrees. After pre-tilting the sample or the devices, the  final alignment for a double tip experiment  can be done with single piezo element. As the 3D alignment is known from SEM images (see Fig.~\ref{fig:DTfinal}), one tilt stage is sufficient, while two stages would further increase the flexibility. Naturally, this will result in a more advanced feedback scheme taking the added tilt stage and both currents into account. Already without a compact tilt stage the devices can be used in single tip operation where the second tip, now out of tunneling, can be used as a gate to locally change the carrier density of dilute electron systems such as semiconductor nanostructures~\cite{gurevich2000a} or underdoped Mott insulators.

\section{Acknowledgments}
We thank R.A.\ Norte for valuable discussions and acknowledge the support from the Kavli Nanolab Delft. This project was supported by the European Research Council (ERC StG Strong-Q and SpinMelt) and by the Netherlands Organisation for Scientific Research (NWO/OCW), as part of the Frontiers of Nanoscience program, as well as through Vidi grants (680-47-536, 680-47-541).

\section*{Conflict of interests}
The authors declare no conflicts of interest.

\section*{Author contributions}
M.P.A.\ and S.G.\ conceived the idea and supervised the project. M.L.\ fabricated the samples, M.L., F.G., K.O., T.B.\ built the STM, M.L., F.G., K.O.\ made the tip holder and M.L.\ and T.B.\ performed the measurements. M.L., M.P.A.\ and S.G.\ wrote the manuscript.

\end{document}